\documentclass[conference]{IEEEtran}
\usepackage{amsmath}
\usepackage{amssymb}
\usepackage[dvips]{graphicx}
\usepackage{layout}

\newtheorem{e_theo}{Theorem}
\newtheorem{e_defin}{Definition}

\newtheorem{e_rema}{Remark}

\newenvironment{e_proof}{\begingroup(\textit{Proof})}{\endgroup}

\newcommand{\vect}[1]{\boldsymbol #1}
\newcommand{\E}{\mathbb{E}}
\def\QED{\hfill$\Box$}	

\begin{document}

\title{Variable-Length Coding with Cost Allowing Non-Vanishing Error Probability}

\author{
 \IEEEauthorblockN{Hideki Yagi}
  \IEEEauthorblockA{
Dept.\ of Computer and Network Engineering\\
University of Electro-Communications \\
Tokyo, Japan\\
Email: h.yagi@uec.ac.jp}
\and
  \IEEEauthorblockN{Ryo Nomura}
  \IEEEauthorblockA{School of Network and Information\\
    Senshu University\\
    Kanagawa, Japan\\
    Email: nomu@isc.senshu-u.ac.jp}}

\maketitle

\begin{abstract}
We derive a general formula of the minimum achievable rate for fixed-to-variable length coding with a regular cost function by allowing the error probability up to a constant $\varepsilon$.
For a fixed-to-variable length code, we call the set of source sequences that can be decoded without error the dominant set of source sequences. For any two regular cost functions, it is revealed that the dominant set of source sequences for a code attaining the minimum achievable rate with a cost function is also the dominant set for a code attaining the minimum achievable rate with the other cost function. We also give a general formula of the second-order minimum achievable rate.
\end{abstract}


\renewcommand\thefootnote{}
\footnotetext{This research is supported by JSPS KAKENHI Grant No.\ 25420357, No.\ 26420371 and No.\ JP16K06340.}
\renewcommand\thefootnote{\arabic{footnote}}

\IEEEpeerreviewmaketitle

\section{Introduction}

For a general source, Han \cite{Han2000} has introduced a notion of ``decoding error'' for variable-length coding and analyzed the minimum average codeword length provided that the decoding error probability vanishes as the source sequence length goes to infinity.
Koga and Yamamoto \cite{Koga-Yamamoto2005} have analyzed the minimum average codeword length for variable-length $\varepsilon$-coding for which the decoding error probability is allowed up to $\varepsilon \in [0,1)$.
For a stationary memoryless source satisfying a certain mild condition, Kostina et al. \cite{KPV2015}
have recently given a single-letter characterization of the optimum second-order codeword length for variable-length $\varepsilon$-codes.

{The problem} of minimizing the average codeword cost with a cost function, which imposes unequal costs for code symbols, has been studied.
This problem, without decoding error, has been introduced by Shannon \cite{Shannon48}. Karp \cite{Karp61} has studied a construction of the optimum prefix code, and Krause \cite{Krause62} has characterized the minimum average codeword cost for stationary memoryless sources.
Han and Uchida \cite{Han-Uchida2000} {have} extended the formula established by \cite{Krause62} to general sources.

In this paper, we introduce the notion of decoding error for variable-length coding with cost.
We first derive finite length upper and lower bounds on the cost rate and  establish a general formula of the minimum achievable cost rate by allowing the error probability up to $\varepsilon$.
We also give a general formula of the second-order minimum achievable rate.
Based on the established second-order coding theorem and the recently obtained result by \cite{KPV2015} (with the uniform cost), a single-letter characterization of  the second-order optimum cost rate is obtained for stationary memoryless sources.

\section{Variable-Length Coding with Cost}

Let $\mathcal{X}$ be a \emph{finite} or \emph{countably infinite} source alphabet.
Let $\vect{X} = \big\{ X^n = \big(X_1^{(n)}, X_2^{(n)}, \ldots, X_n^{(n)} \big) \big\}_{n =1}^\infty$ denote a general source, where $X_i^{(n)}~(i=1,2,\ldots,n)$ takes values in $\mathcal{X}$.
We do not impose any assumptions on $\vect{X}$ such as stationarity or ergodicity.
Let $\mathcal{Y} = \{ 1, \ldots, K\}$ be a code alphabet of size $K$ and let $\mathcal{Y}^*$ denote the set of all finite-length sequences taken from $\mathcal{Y}$. 
We consider a \emph{prefix code} $(\varphi_n, \psi_n)$, where $\varphi_n: \mathcal{X}^n \rightarrow \mathcal{Y}^*$ and $\psi_n:  \mathcal{Y}^* \rightarrow \mathcal{X}^n$ denote an encoder and a decoder, respectively.
Let $\ell (\varphi_n(\vect{x}))$ denote the length of the codeword $\varphi_n(\vect{x})$ for $\vect{x} \in \mathcal{X}^n$.

We now introduce the cost function $c: \mathcal{Y}^* \rightarrow (0,+ \infty)$.
We assume that the cost function can be decomposed for $\vect{y} = (y_1, y_2, \cdots, y_k) \in \mathcal{Y}^k$ as
\begin{align}
c(\vect{y}) = c(y_1) + c(y_2|y_1) + \cdots + c(y_k | y_1^{k-1}), \label{eq:distributed}
\end{align} 
with
\begin{align}
c_{\max} &:= \sup_{ k, y_k, y_1^{k-1}  } c(y_k | y_1^{k-1}) < + \infty, \label{eq:c_max} \\
c_{\min} &:= \inf_{ k, y_k, y_1^{k-1}} c(y_k | y_1^{k-1}) > 0  \label{eq:c_min}
\end{align}
and there exists a unique solution $\alpha = \alpha_c$ of the equation
\begin{align}
&\sum_{y_k \in \mathcal{Y}} K^{ - \alpha c(y_k | y_1^{k-1})} = 1 \label{eq:equation} 
\end{align}
for all  $k = 1, 2, \cdots; y_1^{k-1} \in \mathcal{Y}^{k-1}$.
From \eqref{eq:distributed} and \eqref{eq:equation}, we can easily checked that $\alpha_c$, called the \emph{cost capacity} \cite{Csiszar-Korner2011}, is also the unique solution for the equation
\begin{align}
\sum_{\vect{y} \in \mathcal{Y}^k} K^{ - \alpha c(\vect{y})} = 1~~(\forall k = 1, 2, \cdots).
\end{align}
This class of cost functions, said to be \emph{regular}, was first considered by Han and Kato \cite{Han-Kato97}.
For the prefix code $(\varphi_n, \psi_n)$, we focus on the two performance indices; the \emph{average cost rate}
\begin{align}
\E \left\{ \frac{1}{n} c(\varphi_n(X^n))\right\} = \frac{1}{n}  \sum_{\vect{x} \in \mathcal{X}^n} P_{X^n}(\vect{x})  c(\varphi_n(\vect{x})) 
\end{align}
and the \emph{average error probability} 
\begin{align}
\varepsilon(\varphi_n, \psi_n) := \Pr \{ \psi_n (\varphi_n(X^n)) \neq X^n\}. 
\end{align}
A code of source sequence of length $n$, the average codeword cost $R_n$, and the average error probability $\varepsilon_n$ is called an $(n, R_n, \varepsilon_n)$ code (or simply an $(n, \varepsilon_n)$ code) with cost $c$.

\medskip
\begin{e_rema}
{\rm
Consider a special case where the cost function $c$ satisfies 
\begin{align}
c(y_k | y_1^{k-1}) = 1 ~~(\forall y_k \in \mathcal{Y}; \forall  y_1^{k-1} \in \mathcal{Y}^{k-1}),
\end{align}
where the costs are independent of $y_1^{k-1} \in \mathcal{Y}^{k-1})$.
Then, the cost $c(\varphi_n(\vect{x}))$ of the codeword $\varphi_n(\vect{x})$ is just the codeword length $\ell(\varphi_n(\vect{x}))$.
The average codeword cost  is then the average codeword length, which is often the subject of studies on variable-length source coding.
The codeword cost, which may be asymmetric for $\vect{y} \in \mathcal{Y}^*$,  is a generalized notion of the codeword length. 
\QED}
\end{e_rema}

In this paper, we use the following quantities of a general source $\vect{X}$.
Let $Z$ be a random variable taking values in a (finite or countably infinite) set $\mathcal{Z}$ and let $P_Z$ be its probability measure. Then, for $\delta \in [0,1)$ we define
\begin{align}
\hspace*{-1mm}
G_{[\delta]} (Z) &= \inf_{ \substack{\mathcal{A} \subseteq \mathcal{Z}: \\ \Pr \{ Z \in \mathcal{A} \}  \ge 1- \delta } } \sum_{z \in \mathcal{A}}  P_{Z}(z) \log \frac{\Pr\{ Z\in \mathcal{A} \} }{P_{Z}(z)}, \label{eq:smooth_max_entropy} \\
\hspace*{-1mm}H_{[\delta]} (Z) &= \inf_{ \substack{\mathcal{A} \subseteq \mathcal{Z}: \\ \Pr \{ Z \in \mathcal{A} \}  \ge 1- \delta } } \sum_{z \in \mathcal{A}}  P_{Z}(z) \log \frac{1}{P_{Z}(z)}. \label{eq:smooth_max_entropy2}
\end{align}
In this paper, all logarithms are taken to the base $K$.
Both $G_{[\delta]} (Z)$ and $H_{[\delta]} (Z)$ are nonincreasing functions of $\delta$. 
It obviously holds that $G_{[\delta]} (Z) \le H_{[\delta]} (Z)$ for all $\delta \in [0,1)$.
Based on these quantities, for general source $\vect{X}$ we define 
\begin{align}
H_{[\delta]} (\vect{X}) &= \limsup_{n \rightarrow \infty} \frac{1}{n} H_{[\delta]} (X^n) , \label{eq:smooth_entropy} \\
H_{[\delta]}^* (\vect{X}) &= \liminf_{n \rightarrow \infty} \frac{1}{n} H_{[\delta]} (X^n)  \label{eq:smooth_entropy2}
\end{align}
with a slight abuse of notation.
Obviously $H_{[\delta]}^* (\vect{X}) \le H_{[\delta]} (\vect{X})$, and it is not difficult to verify that
\begin{align}
 H_{[\delta]} (\vect{X}) = \limsup_{n \rightarrow \infty} \frac{1}{n} G_{[\delta]} (X^n) ~~~(\forall \delta \in [0,1)). \label{eq:asymptotic_equivalence}
\end{align}
{It is of use to notice relations among $H_{[\delta]} (\vect{X}), H_{[\delta]}^* (\vect{X})$ and information spectrum quantities \cite{Han2003}.}
Following arguments on $H_{[\delta+\gamma]} (\vect{X}) $ in \cite{Koga-Yamamoto2005, Kuzuoka-Watanabe2015}, we obtain\footnote{A known relation among $H_{[\delta]} (\vect{X})$ and information spectrum quantities is
\begin{align}
\hspace*{-2mm}(1-\delta) \underline{H}(\vect{X}) &  \le  \lim_{\gamma \downarrow 0} H_{[\delta+\gamma]}(\vect{X})  \le  (1-\delta) \overline{H}(\vect{X})~~~~(\delta \in [0,1)),  \nonumber 
\end{align}
where the leftmost inequality is due to Koga and Yamamoto \cite{Koga-Yamamoto2005} whereas the rightmost one is due to Kuzuoka and Watanabe \cite{Kuzuoka-Watanabe2015}.
} 
\begin{align}
(1-\delta) \underline{H}(\vect{X}) &\! \le \! \lim_{\gamma \downarrow 0} H_{[\delta+\gamma]}^* (\vect{X})   \le \! H_{[\delta]}^* (\vect{X})  \! \le \!  (1-\delta) \overline{H}^*(\vect{X}) , \label{eq:information_quantities1} \\
& \hspace*{-10mm} \lim_{\gamma \downarrow 0} H_{[\delta+\gamma]} (\vect{X})  \le \! H_{[\delta]}(\vect{X})  \! \le \!  (1-\delta) \overline{H}(\vect{X}) \label{eq:information_quantities2} 
\end{align}
for every $\delta \in [0,1)$,
where
\begin{align}
&\underline{H}(\vect{X}) \!= \! \sup \left\{a : {\limsup_{n \rightarrow \infty}} \Pr\left\{ \frac{1}{n} \log \frac{1}{P_{X^n}(X^n)} < a \right\} = 0 \right\},  \nonumber  \\
&\overline{H}^* \! (\vect{X}) \! = \! \inf \left\{a : \liminf_{n \rightarrow \infty} \Pr\left\{ \frac{1}{n} \log \frac{1}{P_{X^n}(X^n)} > a \right\} \!  = \! 0 \right\},  \nonumber  \\
&\overline{H} \! (\vect{X}) \! = \! \inf \left\{a : \limsup_{n \rightarrow \infty} \Pr\left\{ \frac{1}{n} \log \frac{1}{P_{X^n}(X^n)} > a \right\} \!  = \! 0 \right\}.  \nonumber
\end{align}
For the proofs of \eqref{eq:information_quantities1} and \eqref{eq:information_quantities2}, see Appendix \ref{Append:prop_information_quantities}.

\section{Finite-Length Analysis}

In this section, we establish finite length lower and upper bounds on the average codeword cost.

\subsection{Converse Bound} \label{sect:converse}

\begin{e_theo}[Converse]\label{theo:converse_bound}
{\rm
Any $(n, R_n, \varepsilon_n)$ prefix code with regular cost $c$ satisfies
 \begin{align}
R_n \ge \frac{G_{[\varepsilon_n]}(X^n) }{\alpha_c n} + \frac{\varepsilon_n c_{\min}}{n}, \label{eq:converse_bound}
\end{align}
where $c_{\min}$ is defined as in \eqref{eq:c_min}.
\QED}
\end{e_theo}

\noindent
(\emph Proof)~~
For an $(n, R_n, \varepsilon_n)$ code $(\varphi_n, \psi_n)$, let
$D_n \subseteq \mathcal{X}^n$ be defined as
 \begin{align}
D_n = \left\{ \vect{x} \in \mathcal{X}^n \Big| \, \psi_n (\varphi_n(\vect{x})) = \vect{x} \right\}.
\end{align}
Then we have $\varepsilon_n = \Pr \{ X^n \in D_n^c\}$ where $D_n^c$ denotes the complement of $D_n$.
It is easily verified that the average codeword cost rate $R_n$ is bounded as
\begin{align}
R_n  &\ge  {\E \left\{ \frac{1}{n} c (\varphi_n(X^n))  \vect{1} \! \left\{ X^n \in D_n \right\}  \right\}}  + \frac{\Pr\{ X^n \in D_n^c \} c_{\min}}{n}, \label{eq:ineq1d} 
\end{align}
where $\vect{1} \{ \cdot\}$ denotes the indicator function.
Defining $q(\vect{y}) = K^{-\alpha_c c(\vect{y})}$ for all $\vect{y} \in \mathcal{Y}^*$,
we have
\begin{align}
\sum_{\vect{x} \in D_n} q(\varphi_n(\vect{x})) \le 1  \label{eq:prefix_cond}
\end{align}
since $\varphi_n$ is one-to-one between $\vect{x} \in D_n$ and $\varphi_n(\vect{x})$.
Then, 
\begin{align}
& \E \left\{ \frac{1}{n} c (\varphi_n(X^n)) \, \Big| X^n \in D_n \right\} \nonumber \\
&= \frac{1}{\alpha_c n} \sum_{\vect{x} \in D_n} \frac{P_{X^n}(\vect{x})}{\Pr\{X^n \in D_n \}}  \log \frac{1}{q(\varphi_n(\vect{x}))} \nonumber \\
&=  \frac{1}{\alpha_c n} \sum_{\vect{x} \in D_n} \frac{P_{X^n}(\vect{x})}{\Pr\{X^n \in D_n \}} \log \frac{P_{X^n}(\vect{x})/\Pr\{X^n \in D_n \}}{q(\varphi_n(\vect{x}))} \nonumber \\
&~~~~~~+ \frac{1}{\alpha_c n} \sum_{\vect{x} \in D_n} \frac{P_{X^n}(\vect{x})}{\Pr\{X^n \in D_n \}}\log \frac{\Pr\{X^n \in D_n \}}{P_{X^n}(\vect{x})} \nonumber \\
& \ge  \frac{1}{\alpha_c n} \sum_{\vect{x} \in D_n} \frac{P_{X^n}(\vect{x})}{\Pr\{X^n \in D_n \}}  \log \frac{\Pr\{X^n \in D_n \}}{P_{X^n}(\vect{x})} \label{eq:ineq1} \\
&\ge \frac{1}{\alpha_c n} \cdot  \frac{G_{[\varepsilon_n]} (X^n) }{\Pr\{X^n \in D_n \}},  \label{eq:ineq2}
\end{align}
where the inequality in \eqref{eq:ineq1} follows due to the log-sum inequality.
Plugging \eqref{eq:ineq2} into \eqref{eq:ineq1d} yields \eqref{eq:converse_bound}.
\QED

\subsection{Achievability Bound} \label{sect:direct_part}

\begin{e_theo}[Achievability] \label{theo:achievability_bound}
{\rm
~~~There exists an $(n, R_n, \varepsilon_n)$ prefix code with regular cost $c$ satisfying 
 \begin{align} 
R_n \le \frac{G_{[\varepsilon_n]}(X^n)}{\alpha_c n}  +\frac{1}{n} \left( \frac{\log 2 + \gamma}{\alpha_c}  + (2 + \varepsilon_n) c_{\max} \right), \label{eq:achievability_bound}
\end{align}
where $\gamma >0$ is an arbitrary constant and $c_{\max}$ is defined as in \eqref{eq:c_max}.
\QED}
\end{e_theo}

\noindent 
(\emph{Proof})~For any $\gamma > 0$ fix a subset $A_n \subseteq \mathcal{X}^n$ such that
\begin{align}
\Pr\{ X^n \in A_n\} \ge 1 - \varepsilon_n
\end{align}
and 
\begin{align}
\sum_{\vect{x} \in A_n } P_{X^n}(\vect{x})  \log \frac{1}{P_{X^n | A_n} (\vect{x})} \le  G_{[\varepsilon_n]} (X^n) + \gamma, \label{eq:set_An}
\end{align}
where we define
\begin{align}
P_{X^n | A_n} (\vect{x})  = \frac{P_{X^n}(\vect{x})}{\Pr\{X^n \in A_n\}}.
\end{align}
Assume that elements of $A_n$ are ordered as $A_n = \{ \vect{x}_1, \vect{x}_2, \cdots \}$.
We use a generalized version of Shannon-Fano-Elias coding with costs (cf.\ \cite{Han-Uchida2000}) for encoding of elements of $A_n$.
 For every $i > 0$ we define  
\begin{align}
P_i &= \sum_{j = 1}^{i-1} P_{X^n | A_n} (\vect{x}_j),  ~~~~Q_i = P_i +  \frac{P_{X^n | A_n} (\vect{x}_i)}{2},
\end{align}
where $P_1 := 0$.
Then, there exists a prefix code $(\tilde{\varphi}_n, \tilde{\psi}_n)$ such that $\varepsilon (\tilde{\varphi}_n, \tilde{\psi}_n) = 0$ and
\begin{align}
K^{- \alpha_c c(\tilde{\varphi}_n(\vect{x}))} > \frac{P_{X^n | A_n} (\vect{x})}{2} K^{-\alpha_c c_{\max}}~~(\forall \vect{x} \in A_n) \label{eq:cost_ineq}
\end{align}
(cf.\ \cite{Han-Uchida2000} and the proof of Theorem \ref{theo:epsilon-achievable-relation} in Section \ref{sect:asymptotic_analysis}). 
We construct a new prefix code $(\varphi_n, \psi_n)$ from $(\tilde{\varphi}_n, \tilde{\psi}_n)$ by setting
\begin{align}
\varphi_n(\vect{x}) = \left\{
\begin{array}{ll}
1 \circ \tilde{\varphi}_n(\vect{x}) & ~\mathrm{if}~\vect{x} \in A_n \\
2 & ~\mathrm{if}~\vect{x} \in A_n^c 
\end{array}
\right.
\end{align}
and 
\begin{align}
\psi_n(\vect{y}) = \left\{
\begin{array}{ll}
\vect{x}_i & ~\mathrm{if}~\vect{y} =\varphi_n(\vect{x}_i) ~\mathrm{with}~\vect{x}_i \in A_n \\
\vect{x}_1 & ~\mathrm{if}~\vect{y} =2 
\end{array}
\right. ,
\end{align}
where $\circ$ denotes concatenation.
Then, it follows from \eqref{eq:cost_ineq} that for all $\vect{x} \in A_n$
\begin{align}
K^{- \alpha_c c(\varphi_n(\vect{x}))} > \frac{P_{X^n | A_n} (\vect{x})}{2} K^{- 2 \alpha_c c_{\max}}. \label{eq:cost_ineq2} 
\end{align}

The decoding error probability is obviously $\varepsilon (\varphi_n, \psi_n) = \Pr\{ X^n \in A_n ^c \} \le \varepsilon_n$.
We evaluate the average cost rate as 
\begin{align}
&\E  \left\{ \frac{1}{n} c (\varphi_n(X^n)) \right\} \nonumber \\
&~\le \Pr\{ X^n \in A_n \} \E \left\{ \frac{1}{n} c (\varphi_n(X^n)) \, \Big| X^n \in A_n \right\} + \varepsilon_n \frac{c(2)}{n}. \label{eq:ineq3}
\end{align}
In view of \eqref{eq:cost_ineq2}, the first term is evaluated as
\begin{align}
 \Pr & \{ X^n \in A_n \} \E \left\{ \frac{1}{n} c (\varphi_n(X^n)) \, \Big| X^n \in A_n \right\}  \nonumber \\
 &\le \frac{1}{\alpha_c n} \Pr\{ X^n \in A_n \} \E \left\{  \log \frac{1}{P_{X^n| A_n}(X^n)} \, \Big| X^n \in A_n \right\}  \nonumber \\
 &~~+ \frac{\log 2}{\alpha_c n} + \frac{2 c_{\max}}{n} \nonumber \\
&\le \frac{G_{[\varepsilon_n]}(X^n)}{\alpha_c n}  +  \frac{\log 2 + \gamma}{\alpha_c n} + \frac{2 c_{\max}}{n}, \label{eq:ineq4}
\end{align}
where we have used \eqref{eq:set_An} for the last inequality.
Plugging \eqref{eq:ineq4} into \eqref{eq:ineq3} yields \eqref{eq:achievability_bound}.
\QED

\section{Asymptotic Analysis} \label{sect:asymptotic_analysis}

\smallskip

\subsection{Definitions}

We define the $\varepsilon$-achievable cost rates as follows:
\begin{e_defin}[Type-I $\varepsilon$-Achievable Cost Rate]
{\rm
For $\varepsilon \in (0,1)$, a cost rate $R \ge 0$ is said to be \emph{type-I $\varepsilon$-achievable with cost $c$} if there exists a sequence of 
$(n, \varepsilon_n)$ codes satisfying
\begin{align}
 &\limsup_{n \rightarrow \infty} \E \left\{ \frac{1}{n} c(\varphi_n(X^n))\right\} \le R, \label{eq:cost-rate_cond}  \\ 
 &\varepsilon_n \le \varepsilon~~~(\forall n > n_0) \label{eq:typeI_error_prob_cond}   .
\end{align}
The infimum of all type-I $\varepsilon$-achievable cost rates with cost $c$ is denoted by $\mathcal{R}_c^{(\mathrm{I})} (\varepsilon| \vect{X})$.
Also, $R \ge 0$ is said to be \emph{type-I optimistically $\varepsilon$-achievable with cost} $c$ if there exists a sequence of $(n, \varepsilon_n)$ codes satisfying
\begin{align}
&\liminf_{n \rightarrow \infty} \E \left\{ \frac{1}{n} c(\varphi_n(X^n))\right\} \le R, \nonumber \\
&\varepsilon_n \le \varepsilon~~(\forall n > n_0). \label{eq:typeI_opt_error_prob_cond}
\end{align}
The infimum of all optimistically $\varepsilon$-achievable cost rates with cost $c$ is  denoted by $\mathcal{R}_c^{(\mathrm{I})*}  (\varepsilon| \vect{X})$.
\QED}
\end{e_defin}

The following definition gives a right-continuous version of the infimum $\varepsilon$-achievable cost rate, which is a generalized notion of \emph{weak achievability} for variable-length codes (cf. Han \cite{Han2000}, Koga and Yamamoto \cite{Koga-Yamamoto2005}). 

\begin{e_defin}[Type-II $\varepsilon$-Achievable Cost Rate]
{\rm
For $\varepsilon \in [0,1)$, a cost rate $R \ge 0$ is said to be \emph{type-II $\varepsilon$-achievable with cost $c$} if there exists a sequence of 
$(n, \varepsilon_n)$ codes satisfying \eqref{eq:cost-rate_cond} and
\begin{align}
\limsup_{n \rightarrow \infty} \varepsilon_n \le \varepsilon. \label{eq:typeII_error_prob_cond}
\end{align}
The infimum of all type-II  $\varepsilon$-achievable cost rates with cost $c$ is denoted by $\mathcal{R}_c^{(\mathrm{II})}  (\varepsilon| \vect{X})$.
\QED}
\end{e_defin}

\begin{e_rema} \label{rema:type-Iand-II}
{\rm
It is easily shown that we have 
\begin{align}
 \mathcal{R}_c^{(\mathrm{II})}  (\varepsilon| \vect{X}) = \lim_{\gamma \downarrow 0} \mathcal{R}_c^{(\mathrm{I})}  (\varepsilon+ \gamma| \vect{X})~~(\forall \varepsilon \in [0,1)). 
\end{align}
We have the analogous relation for optimistically $\varepsilon$-achievable cost rates.
This means that it suffices to establish a formula for type-I $\varepsilon$-achievable cost rates, so we shall consider only the  type-I achievability.
\QED}
\end{e_rema}

\subsection{First-Order Coding Theorem}

Now, we establish the general formula for the type-I $\varepsilon$-achievable cost rates.

\begin{e_theo}[Type-I $\varepsilon$-Achievable Cost Rate] \label{theo:typeI-epsilon-achievable-rate}
{\rm
For every $\varepsilon \in (0,1)$, any general source $\vect{X}$ satisfies 
\begin{align}
\mathcal{R}_c^{(\mathrm{I})} (\varepsilon| \vect{X}) &= \frac{H_{[\varepsilon]}(\vect{X})}{\alpha_c} = \limsup_{n \rightarrow \infty} \frac{ H_{[\varepsilon]} (X^n)}{\alpha_c n} , \label{eq:typeI_min_rate} \\
\mathcal{R}_c^{(\mathrm{I})*}  (\varepsilon| \vect{X}) &= \frac{H_{[\varepsilon]}^*(\vect{X})}{\alpha_c} = \liminf_{n \rightarrow \infty} \frac{H_{[\varepsilon]} (X^n)}{\alpha_c n} . \label{eq:typeI_opt_min_rate} 
\end{align}
\QED}
\end{e_theo}

\begin{e_rema} \label{rema:previous_formulas}
{\rm
Formulas \eqref{eq:typeI_min_rate} and \eqref{eq:typeI_opt_min_rate} are established for the first time even when $c = \ell$ (i.e., $\alpha_c =1$).
Based on Remark \ref{rema:type-Iand-II}, formulas \eqref{eq:typeI_min_rate} and \eqref{eq:typeI_opt_min_rate} lead to the general formulas for the type-II achievable rate cost rates, which generalize formulas for the $\varepsilon$-achievable rate with uniform cost $c = \ell$ given by \cite{Han2000} and \cite{Koga-Yamamoto2005} and the general formula for the achievable rate with regular cost $c$ and $\varepsilon = 0$ given by \cite{Han-Uchida2000}.}
\end{e_rema}

\medskip
\noindent 
\emph{Proof of Converse Part:}~
We shall show the formula for $\mathcal{R}_c^{(\mathrm{I})} (\varepsilon|\vect{X}) $.
The formula for $\mathcal{R}_c^{(\mathrm{I})*}  (\varepsilon|\vect{X}) $ can be proven in a similar way.

Let $R \ge  0$ be type-I $\varepsilon$-achievable with cost $c$. Then, by definition, there exists a sequence of $(n, R_n, \varepsilon_n)$ codes $(\varphi_n, \psi_n)$ satisfying \eqref{eq:cost-rate_cond}  and \eqref{eq:typeI_error_prob_cond}.
Theorem \ref{theo:converse_bound} assures that for such codes we have for all $n > 0$,
\begin{align}
R_n = \frac{1}{n} \E \{ c(\varphi_n(X^n))\}  \ge \frac{ G_{[\varepsilon_n]} (X^n)}{\alpha_c n}. \label{eq:rate_bound1}
\end{align}
It follows from \eqref{eq:typeI_error_prob_cond} that
\begin{align}
\frac{1}{n} \E \{ c(\varphi_n(X^n))\}  \ge  \frac{ G_{[\varepsilon]} (X^n)}{\alpha_c n}~~~(\forall n > n_0)
\end{align}
because $G_{[\delta]}(X^n)$ is a nonincreasing function in $\delta$.
Thus,
\begin{align}
R \ge \limsup_{n \rightarrow \infty} \frac{1}{n} \E \{ c(\varphi_n(X^n))\}  &\ge \frac{H_{[\varepsilon]} (\vect{X})}{\alpha_c}, \nonumber
\end{align}
where we have used the relation \eqref{eq:asymptotic_equivalence}.
\QED

\medskip
\noindent 
\emph{Proof of Direct Part:}~
We shall show the formula for $\mathcal{R}_c^{(\mathrm{I})} (\varepsilon|\vect{X}) $.
The formula for $\mathcal{R}_c^{(\mathrm{I})*}  (\varepsilon|\vect{X}) $ can be proven in a similar way.

Let $\{ \varepsilon_n \}_{n=1}^\infty$ be a sequence such that $\varepsilon_n > 0$ and
\begin{align}
\varepsilon_n = \varepsilon ~~(\forall n > n_0) \label{eq:type_I_error_cond2}
\end{align}
Theorem \ref{theo:achievability_bound} assures that for any $\gamma > 0$ there exists an $(n, R_n ,\varepsilon_n)$ code $(\varphi_n, \psi_n)$ such that 
\begin{align}
R_n   = \frac{1}{n} \E \{ c(\varphi_n(X^n))\} \le \frac{ G_{[\varepsilon_n]} (X^n)}{\alpha_c n} + \gamma~~~(\forall n > n_1). \nonumber 
\end{align}
It follows from \eqref{eq:type_I_error_cond2} that
\begin{align}
\limsup_{n \rightarrow \infty } \frac{1}{n} \E \{ c(\varphi_n(X^n))\} 
&\le \limsup_{n \rightarrow \infty } \frac{ G_{[\varepsilon]} (X^n)}{\alpha_c n} + \gamma. \nonumber
\end{align}
Since $\gamma >0$ is an arbitrary constant, this inequality and the relation \eqref{eq:asymptotic_equivalence} mean that
$\mathcal{R}_c^{(\mathrm{I})} (\varepsilon|\vect{X}) \le H_{[\varepsilon]} (\vect{X})/ \alpha_c$. 
\QED 

\subsection{Relation Between Achievable Rates with Different Costs} \label{sect:epsilon-achievable-relation_proof}

Now, we turn to discussing a relationship between the $\varepsilon$-achievable cost rates under two different cost functions. 
Although the following theorem is an immediate consequence of Theorem \ref{theo:typeI-epsilon-achievable-rate}, we describe an alternative proof which leads to an observation on the structure of optimal codes with distinct cost functions (cf. Remark \ref{rema:dominant_set}). 
\begin{e_theo} \label{theo:epsilon-achievable-relation}
{\rm
Let $c, c'$ be regular cost functions and let $\alpha_c$ and $\alpha_{c'}$ denote the unique solution of  equation \eqref{eq:equation} for each cost function.
Then, for every $\varepsilon \in (0,1)$ we have
\begin{align}
\alpha_c \mathcal{R}_c^{(\mathrm{I})} (\varepsilon| \vect{X}) &= \alpha_{c'} \mathcal{R}_{c'}^{(\mathrm{I})} (\varepsilon| \vect{X}), \\
\alpha_c \mathcal{R}_c^{(\mathrm{I})*} (\varepsilon| \vect{X}) &= \alpha_{c'} \mathcal{R}_{c'}^{(\mathrm{I})*} (\varepsilon| \vect{X}).
\end{align}
}
\end{e_theo}
\begin{e_proof}
It suffices to show the following claims:
\begin{itemize}
\item[(i)]
If $R$ is type-I (resp. type-II) $\varepsilon$-achievable with cost $c$, then $\frac{\alpha_c}{\alpha_{c'}} \cdot R$ is type-I (resp. type-II) $\varepsilon$-achievable with cost $c'$.
\item[(ii)] If $R$ is type-I (resp. type-II) optimistically $\varepsilon$-achievable with cost $c$, then $\frac{\alpha_c}{\alpha_{c'}} \cdot R$ is type-I (resp. type-II) optimistically $\varepsilon$-achievable with cost $c'$.
\end{itemize}
These claims may be proven by applying \cite[Lemma 1]{Uyematsu-Kawakami2000} twice. 
Here, we give a slightly more direct proof.

For a type-I $\varepsilon$-achievable cost rate $R$ with cost $c$, there exists a prefix code $(\varphi_n, \psi_n)$ satisfying \eqref{eq:cost-rate_cond} and \eqref{eq:typeI_error_prob_cond}.
Set 
\begin{align}
D_n = \left\{ \vect{x} \in \mathcal{X}^n : ~ \psi_n(\varphi_n(\vect{x})) = \vect{x} \right\}.
\end{align}
By definition, we have $\varepsilon_n = \Pr\{ X^n \in D_n^c\}$. Then, similarly to the derivation of \eqref{eq:ineq1}, we have 
\begin{align}
\hspace*{-3mm}\E  \left\{ \frac{c (\varphi_n(X^n))}{n}  \right\} &\ge  { \E \left\{ \frac{c (\varphi_n(X^n)) }{n}  \vect{1} \! \left\{ X^n \in D_n \right\}  \right\}}
\nonumber \\
&\ge \frac{1}{\alpha_c n} \sum_{\vect{x} \in D_n} P_{X^n}(\vect{x})  \log \frac{1}{P_{X^n|D_n} (\vect{x})}, \label{eq:ineq1b}
\end{align}
where we define
\begin{align}
P_{X^n|D_n} (\vect{x}) = \frac{P_{X^n} (\vect{x})}{\Pr\{ X^n \in D_n \}}~~(\forall \vect{x} \in \mathcal{X}^n).
\end{align}
We use a generalized version of Shannon-Fano-Elias coding with costs (cf.\ \cite{Han-Uchida2000}).
Assume that the elements of $D_n$ are indexed as $D_n = \{ \vect{x}_1, \vect{x}_2, \cdots \}$.
We define
\begin{align}
P_i &:= \sum_{j=1}^{i-1} P_{X^n|D_n} (\vect{x}_j), ~~~~Q_i :=  P_i + \frac{ P_{X^n|D_n} (\vect{x}_i)}{2}
\end{align}
for all $i = 1, 2, \cdots$, where $P_1 := 0.$
For the cost function $c'$ with $q(\vect{y}) = K^{-\alpha_{c'} c'(\vect{y})}~(\forall \vect{y} \in \mathcal{Y}^*)$,
we also define 
\begin{align}
I(\vect{y}) &= [\beta(\vect{y}), \gamma(\vect{y})), \\
\beta(\vect{y}) &= \sum_{\vect{y}' : \vect{y}' \prec \vect{y}} q(\vect{y}')~~\mathrm{and}~~
\gamma(\vect{y}) = \beta(\vect{y}) + q(\vect{y}), 
\end{align}
where $\prec$ denotes the lexicographic order on the set $\mathcal{Y}^{\ell(\vect{y})}$.
Now, to each $\vect{x}_i$ we assign $\vect{y}_i$ as
\begin{align}
\vect{y}_i = \arg \min_{\vect{y} \in \mathcal{K}_i} \ell(\vect{y}),
\end{align}
where $\mathcal{K}_i$ is the set of $\vect{y} \in \mathcal{Y}^*$ such that $I(\vect{y})$ includes $Q_i$ but neither $P_i$ nor $P_{i+1}$. 
Then, it holds that $I(\vect{y}_i) \subset (P_i, P_{i+1})$ and intervals $I(\vect{y}_1), I(\vect{y}_2), \cdots$ are disjoint, implying that $\{ \vect{y}_1, \vect{y}_2, \cdots \}$ forms a prefix code.
We arrange a new encoder $\varphi_n' : \mathcal{X}^n \rightarrow \mathcal{Y}^*$ as
\begin{align}
\varphi_n'(\vect{x}_i) = \left\{
\begin{array}{ll}
1 \circ \vect{y}_i & \mathrm{if}~\vect{x}_i \in D_n\\
2 & \mathrm{if}~\vect{x}_i \not\in D_n,
\end{array}
\right.
\end{align}
where $\circ$ denotes concatenation.
The decoder $\psi_n'$ is such that $\psi_n'(\varphi_n'(\vect{x}_i)) = \vect{x}_i$ for all $\vect{x}_i \in D_n$.
Therefore, the decoding error probability does not change and the code $(\varphi_n', \psi_n')$ satisfies \eqref{eq:typeI_error_prob_cond}.

Now, for each $\vect{y} = (y_1, y_2, \ldots, y_l)$, where $l = \ell(\vect{y})$, set $\overline{\vect{y}}_i =(y_1, y_2, \ldots, y_{l -1}) $.
Then, by definition, $I(\vect{y}_i) \subset I(\overline{\vect{y}}_i)$ and $P_i \in I(\overline{\vect{y}}_i)$ or $P_{i+1} \in I(\overline{\vect{y}}_i)$.
This means that the width $|I(\overline{\vect{y}}_i)|$ of the interval $I(\overline{\vect{y}}_i)$ is larger than $P_{X^n|D_n}(\vect{x}_i)/2$, so that
\begin{align}
|I(\overline{\vect{y}}_i)| = K^{- \alpha_{c'} c'(\overline{\vect{y}}_i)} > \frac{P_{X^n|D_n}(\vect{x}_i)}{2}.
\end{align}
Since 
\begin{align}
c'(\varphi_n'(\vect{x}_i)) \le c'(\vect{y}_i) + c'_{\max} \le  c'(\overline{\vect{y}}_i) + 2 c'_{\max}~~(\forall \vect{x}_i \in D_n), \nonumber
\end{align}
we obtain 
\begin{align}
c'(\varphi_n'(\vect{x}_i)) \!\le \! \left\{ \!
\begin{array}{ll}
\!\! \frac{- \log P_{X^n|D_n}(\vect{x}_i)}{\alpha_{c'}} + \frac{ \log 2}{\alpha_{c'}} +  2c'_{\max} \!  & \!  \mathrm{if}~\vect{x}_i \in D_n\\
\!\! c'_{\max} \! & \! \mathrm{if}~\vect{x}_i \not\in D_n.
\end{array}
\right. \nonumber
\end{align}
Then, we obtain
\begin{align}
&\limsup_{n \rightarrow \infty} \E  \left\{ \frac{1}{n} c' (\varphi_n'(X^n)) \right\} \nonumber \\
&~~\le \limsup_{n \rightarrow \infty} \frac{1}{\alpha_{c'} n} \sum_{\vect{x} \in D_n} P_{X^n}(\vect{x}) \cdot \log \frac{1}{P_{X^n|D_n} (\vect{x})} \nonumber \\
&~~ \le \frac{\alpha_c}{\alpha_{c'}} \limsup_{n \rightarrow \infty} \E  \left\{ \frac{1}{n} c (\varphi_n(X^n)) \right\} \le \frac{\alpha_c}{\alpha_{c'}} \cdot R,
\end{align}
where we have used \eqref{eq:cost-rate_cond} and \eqref{eq:ineq1b}.
Thus, the proof of claim (i) is completed. 
Claim (ii) can be proven similarly.
\QED
\end{e_proof}

\begin{e_rema} \label{rema:dominant_set}
{\rm
In the foregoing proof, a good $(n, \varepsilon_n)$ code for cost $c'$ is obtained from a good $(n, \varepsilon_n)$ code for cost $c$ without changing the \emph{dominant set} $D_n$, which is the set of source sequences that can be decoded without error.
This means that for any two regular cost functions, the dominant set for a code that attains the infimum $\varepsilon$-achievable cost rate with a cost function is also the dominant set for a code attaining the  infimum  $\varepsilon$-achievable cost rate with the other cost function.  
\QED}
\end{e_rema}

\section{Optimum Second-Order Cost Rate}

\subsection{Definitions}

We define the second-order achievable cost rates as follows: 

\begin{e_defin}[Type-I $(\varepsilon, R)$-Achievable Cost Rate]
{\rm
For $\varepsilon \in (0,1)$ and $R \ge 0$, $L $ is said to be second-order \emph{type-I $(\varepsilon, R)$-achievable with cost $c$} if there exists a sequence of 
$(n, \varepsilon_n)$ codes satisfying
\begin{align}
 &\limsup_{n \rightarrow \infty} \frac{1}{\sqrt{n}} \big( \E \left\{ c(\varphi_n(X^n))\right\}  - n R \big) \le L, \label{eq:2nd_cost-rate_cond}  \\ 
 &\varepsilon_n \le \varepsilon~~~(\forall n > n_0) \label{eq:2nd_typeI_error_prob_cond}   .
\end{align}
The infimum of all type-I $(\varepsilon,R)$-achievable cost rates with cost $c$ is denoted by $\mathcal{L}_c^{(\mathrm{I})} (\varepsilon, R| \vect{X})$.
Also, $L $ is said to be second-order \emph{type-I optimistically $(\varepsilon, R)$-achievable with cost} $c$ if there exists a sequence of $(n, \varepsilon_n)$ codes satisfying
\begin{align}
&\liminf_{n \rightarrow \infty} \frac{1}{\sqrt{n}} \big( \E \left\{ c(\varphi_n(X^n))\right\}  - n R \big)   \le L, \nonumber \\
& \varepsilon_n \le \varepsilon ~~~~(\forall n > n_0). \label{eq:2nd_typeI_opt_error_prob_cond}
\end{align}
The infimum of all type-I optimistically $(\varepsilon,R)$-achievable cost rates with cost $c$ is  denoted by $\mathcal{L}_c^{(\mathrm{I})*}  (\varepsilon, R| \vect{X})$.
\QED}
\end{e_defin}

\begin{e_rema} \label{rema:2nd_type-Iand-II}
{\rm
Similarly to the first-order cost rates, we can also define a right-continuous version of the infimum $(\varepsilon, R)$-achievable rate (called type-II $(\varepsilon, R)$-achievable cost rate), denoted by $\mathcal{L}_c^{(\mathrm{II})} (\varepsilon, R| \vect{X})$, by replacing \eqref{eq:2nd_typeI_error_prob_cond} with
\begin{align}
\limsup_{n \rightarrow \infty} \varepsilon_n \le \varepsilon.
\end{align}  
Then, for $\varepsilon \in [0,1)$ we have 
\begin{align}
 \mathcal{L}_c^{(\mathrm{II})}  (\varepsilon, R| \vect{X}) = \lim_{\gamma \downarrow 0} \mathcal{L}_c^{(\mathrm{I})}  (\varepsilon+ \gamma, R| \vect{X}). 
\end{align}
\QED}
\end{e_rema}

\subsection{Second-Order Coding Theorem}

We establish the second-order coding theorem, which is a counterpart of Theorem \ref{theo:typeI-epsilon-achievable-rate} of the first-order. 
\begin{e_theo}[Type-I $(\varepsilon, R)$-Achievable Cost Rate] \label{theo:2nd_typeI-epsilon-achievable-rate}
{\rm
For every $\varepsilon \in (0,1)$ and $R \ge 0$, any general source $\vect{X}$ satisfies 
\begin{align}
\mathcal{L}_c^{(\mathrm{I})} (\varepsilon, R| \vect{X}) & = \limsup_{n \rightarrow \infty} \frac{1}{ \sqrt{n}} \left(\frac{H_{[\varepsilon]} (X^n)}{\alpha_c} - n R \right) , \label{eq:2nd_typeI_min_rate} \\
\mathcal{L}_c^{(\mathrm{I})*}  (\varepsilon, R| \vect{X}) &=\liminf_{n \rightarrow \infty} \frac{1}{ \sqrt{n}} \left(\frac{H_{[\varepsilon]} (X^n)}{\alpha_c} - n R \right) . \label{eq:2nd_typeI_opt_min_rate} 
\end{align}
}
\end{e_theo}
(\emph{Proof})  ~Using the relation
\begin{align}
\limsup_{n \rightarrow \infty} \frac{1}{ \sqrt{n}} H_{[\varepsilon]} (X^n) = \limsup_{n \rightarrow \infty} \frac{1}{ \sqrt{n}} G_{[\varepsilon]} (X^n),
\end{align}
we can prove the theorem similarly to Theorem \ref{theo:typeI-epsilon-achievable-rate}.
\QED

\begin{e_rema}
{\rm
For the case where $c = \ell$, we have the following immediate consequence of Theorem \ref{theo:2nd_typeI-epsilon-achievable-rate}:
for every $\varepsilon \in (0,1)$ and $R \ge 0$, any general source $\vect{X}$ satisfies
\begin{align}
\mathcal{L}_\ell^{(\mathrm{I})} (\varepsilon, R| \vect{X}) & = \limsup_{n \rightarrow \infty} \frac{1}{ \sqrt{n}} (H_{[\varepsilon]} (X^n) - n R) . \label{eq:2nd_typeI_min_rate2}
\end{align}
Thus, we have
\begin{align}
\alpha_c \mathcal{L}_c^{(\mathrm{I})} (\varepsilon, R| \vect{X}) = \mathcal{L}_\ell^{(\mathrm{I})} (\varepsilon, \alpha_c R| \vect{X})\label{eq:2nd_typeI_relation}
\end{align}
for any regular cost function $c$.
\QED}
\end{e_rema}

In the case where  $c = \ell$ and the source $\vect{X}$ is stationary and memoryless  with the finite third absolute moment of $\log \frac{1}{P_X(X)}$, Kostina et al.\ \cite{KPV2015} has recently given a single-letter characterization of $\mathcal{L}_\ell^{(\mathrm{I})} (\varepsilon, R| \vect{X})$ with $R = H_{[\varepsilon]}(\vect{X})$ as
\begin{align}
\mathcal{L}_\ell^{(\mathrm{I})} (\varepsilon, R| \vect{X}) = - \sqrt{\frac{V(X)}{2 \pi}} e^{-\frac{(Q^{-1}(\varepsilon))^2}{2}},
\end{align}
where $V(X)$ denotes the variance of $\log \frac{1}{P_X(X)}$ (varentropy) and $Q^{-1}$ is the inverse of the complementary cumulative distribution function of the standard Gaussian distribution.
Notice that $R = H_{[\varepsilon]}(\vect{X}) = (1-\varepsilon)H(X)$ in this case \cite{Koga-Yamamoto2005}, where $H(X)$ is the entropy of the source. 
Now, let us consider the case where the cost function is \emph{additive} \cite{Csiszar-Korner2011}.
In view of the relation \eqref{eq:2nd_typeI_relation}, we can also obtain a single-letter characterization 
\begin{align}
\mathcal{L}_c^{(\mathrm{I})} (\varepsilon, R| \vect{X}) & = - \frac{1}{\alpha_c} \sqrt{\frac{ V(X)}{2 \pi}} e^{-\frac{(Q^{-1}(\varepsilon))^2}{2}} ,
\end{align}
where the first-order cost rate is $R =H_{[\varepsilon]}(\vect{X})/\alpha_c$.
As is observed in \cite{KPV2015}, it is of interest to see that the optimum second-order $(\varepsilon, R)$-achievable cost rate is always negative, and allowing the decoding error up to $\varepsilon$ is beneficial for both the first- and second-order cost rates.

\appendices

\section{Proofs of Equations \eqref{eq:information_quantities1}  and \eqref{eq:information_quantities2} } \label{Append:prop_information_quantities}

We shall prove (i) $(1-\delta) \underline{H}(\vect{X}) \le \lim_{\gamma \downarrow 0} H_{[\delta+\gamma]}^* (\vect{X})$, (ii) $H_{[\delta]}^* (\vect{X}) \le (1-\delta) \overline{H}^*(\vect{X})$, and (iii) $H_{[\delta]} (\vect{X}) \le (1-\delta) \overline{H}(\vect{X})$ because other inequalities are trivial.

\smallskip
{(i) \emph{Proof of $(1-\delta) \underline{H}(\vect{X}) \le \lim_{\gamma \downarrow 0} H_{[\delta+\gamma]}^* (\vect{X})$}}:~~This inequality can be proven similarly to \cite[Theorem 4]{Koga-Yamamoto2005} and \cite[Theorem 3]{Kuzuoka-Watanabe2015}, which show $(1-\delta) \underline{H}(\vect{X}) \le \lim_{\gamma \downarrow 0} H_{[\delta + \gamma]} (\vect{X})$.
 We describe the whole proof for readers' convenience.

Fix $\gamma>0$ and $\eta > 0$ arbitrarily. For all $n=1,2,\cdots$, we choose a subset $A_n \subseteq \mathcal{X}^n$ such that
\begin{align}
\hspace*{-3mm} \Pr \left\{ X^n \in A_n \right\} &\ge 1 - \delta - \gamma, \\
\hspace*{-3mm} \frac{1}{n} \sum_{\vect{x} \in A_n } \! P_{X^n}(\vect{x}) \log \frac{1}{P_{X^n}(\vect{x})}  &\le \frac{1}{n} H_{[\delta+\gamma]}(X^n) + \eta. \label{eq:set:An}
\end{align}
Set
\begin{align}
T_n = \left\{ \vect{x} \in \mathcal{X}^n : \frac{1}{n} \log \frac{1}{P_{X^n}(\vect{x})} \ge \underline{H}(\vect{X}) - \eta \right\}.
\end{align}
Then, for sufficiently large $n$ we have
\begin{align}
\Pr \left\{ X^n \in A_n \cap T_n  \right\} &\ge \Pr \left\{ X^n \in A_n \right\} - \Pr \left\{ X^n \in T_n^c \right\} \nonumber \\
&\ge  1- \delta - 2 \gamma,
\end{align}
where the last inequality is due to the definition  of $\underline{H}(\vect{X})$.
We obtain
\begin{align}
&\frac{1}{n} \sum_{\vect{x} \in A_n} P_{X^n}(\vect{x}) \log \frac{1}{P_{X^n}(\vect{x})}  \nonumber \\
&~~\ge \frac{1}{n} \sum_{\vect{x} \in A_n \cap T_n}  P_{X^n}(\vect{x}) \log \frac{1}{P_{X^n}(\vect{x})} \nonumber \\
&~~ \ge \Pr \{ X^n \in A_n \cap T_n\} (\underline{H}(\vect{X}) - \eta) \nonumber \\
&~~ \ge ( 1- \delta - 2 \gamma)(\underline{H}(\vect{X}) - \eta).
\end{align}
It follows from \eqref{eq:set:An} that
\begin{align}
H_{[\delta+\gamma]}^*(\vect{X}) &\ge \liminf_{n \rightarrow \infty}  \frac{1}{n} \sum_{\vect{x} \in A_n} \log \frac{1}{P_{X^n}(\vect{x})} - \eta \nonumber \\
& \ge  ( 1- \delta - 2 \gamma)(\underline{H}(\vect{X}) - \eta) - \eta.
\end{align}
Since $\eta>0$ is arbitrary, we obtain the inequality $(1-\delta - 2 \gamma) \underline{H}(\vect{X}) \le H_{[\delta+\gamma]}^* (\vect{X})$.
By taking $\lim_{\gamma \downarrow 0}$, we have proven the inequality $(1-\delta) \underline{H}(\vect{X}) \le \lim_{\gamma \downarrow 0} H_{[\delta+\gamma]}^* (\vect{X})$.

\smallskip
{(ii) \emph{Proof of $H_{[\delta]}^* (\vect{X}) \le (1-\delta) \overline{H}^*(\vect{X})$}}:~~
Set
\begin{align}
\hspace*{-3mm} S_n = \left\{ \vect{x} \in \mathcal{X}^n : \frac{1}{n} \log \frac{1}{P_{X^n}(\vect{x})} \le \overline{H}^*(\vect{X}) + \gamma \right\},
\end{align}
where $\gamma >0$ is an arbitrary constant.
In view of the equation
\begin{align}
\liminf_{n \rightarrow \infty} \Pr \left\{ X^n \in S_n^c \right\} = 0,
\end{align}
let $n_1 < n_2 < \cdots$ denote an increasing sequence such that
\begin{align}
\lim_{i \rightarrow \infty} \Pr \left\{ X^{n_i} \in S_{n_i}^c \right\} = 0. \label{eq:sub_seq_n}
\end{align}
We fix any $\delta' \in (0,\delta)$. For all $i=1,2,\cdots$, we choose a subset $B_{n_i} \subseteq \mathcal{X}^{n_i}$ such that
\begin{align}
\hspace*{-4mm} 1 - \delta' &\le \Pr \left\{ X^{n_i} \in B_{n_i} \right\}, \label{eq:set:Bn1} \\
\hspace*{-4mm} 1 - \delta' &\ge \Pr \left\{ X^{n_i} \in \Gamma \right\}   ~~~(\forall \Gamma \subset B_{n_i} ~\mathrm{s.t.}~ \Gamma \neq B_{n_i}). \label{eq:set:Bn2}
\end{align}
Notice that we can always choose such $B_{n_i} \subseteq \mathcal{X}^{n_i}$, for example, by successively inserting $\vect{x} \in \mathcal{X}^{n_i}$ to  $B_{n_i}$ in the decreasing order of $P_{X^{n_i}}(\vect{x})$ and stop this procedure once \eqref{eq:set:Bn1} is satisfied.
From \eqref{eq:sub_seq_n} and \eqref{eq:set:Bn1} we have
\begin{align}
&\Pr \left\{ X^{n_i} \in B_{n_i} \cap S_{n_i}  \right\} \nonumber \\
&~~~~\ge \Pr \left\{ X^{n_i} \in B_{n_i} \right\} - \Pr \left\{ X^{n_i} \in S_{n_i}^c \right\} \nonumber \\
&~~~~\ge  1- \delta' - \gamma ~~~~(\forall i > i_0). \label{eq:ineq0}
\end{align}
On the other hand, fixing an arbitrary $\vect{x}_0 \in B_{n_i}$ with $p_0 := P_{X^n}(\vect{x}_0)$ and setting $\tilde{B}_{n_i} = B_{n_i} \setminus \{\vect{x}_0\}$, we have 
\begin{align}
& \frac{1}{n_i} \sum_{\vect{x} \in B_{n_i} \cap S_{n_i}} P_{X^{n_i}}(\vect{x}) \log \frac{1}{P_{X^{n_i}}(\vect{x})} \nonumber \\
&~~\le \frac{1}{n_i} \sum_{\vect{x} \in \tilde{B}_{n_i} \cap S_{n_i}} P_{X^{n_i}}(\vect{x}) \log \frac{1}{P_{X^{n_i}}(\vect{x})}  + \frac{p_0}{n_i} \log \frac{1}{p_0} \nonumber \\
&~~ \le \Pr \{ X^{n_i} \in \tilde{B}_{n_i} \cap S_{n_i}\} (\overline{H}^*(\vect{X}) + \gamma) + \frac{p_0}{n_i} \log \frac{1}{p_0} \nonumber \\
&~~ \le ( 1- \delta')(\overline{H}^*(\vect{X}) + \gamma) + \frac{\log e}{n_ie},  \label{eq:ineq0b}
\end{align}
where the second inequality is due to the definition of $S_{n}$ and the last inequality is due to \eqref{eq:set:Bn2} and  $p_0 \log p_0 \ge - \frac{\log e}{e}$ for $p_0 \in [0,1]$.
It follows from \eqref{eq:ineq0} that
 \begin{align}
\hspace*{-3mm} \frac{1}{n_i} H_{[\delta'+\gamma]}(X^{n_i})\le  \frac{1}{n_i} \sum_{\vect{x} \in B_{n_i} \cap S_{n_i}} P_{X^{n_i}}(\vect{x}) \log \frac{1}{P_{X^{n_i}}(\vect{x})} \nonumber 
\end{align}
and thus from \eqref{eq:ineq0b} that
\begin{align}
\frac{1}{n_i} H_{[\delta'+\gamma]}(X^{n_i}) \le   ( 1- \delta')(\overline{H}^*(\vect{X}) + \gamma) + \frac{\log e}{n_ie} \nonumber 
\end{align}
for all $i > i_0$, 
which leads to
\begin{align}
&\liminf_{n \rightarrow \infty} \frac{1}{n} H_{[\delta'+\gamma]}(X^{n}) \nonumber \\
&~~\le \liminf_{i \rightarrow \infty} \frac{1}{n_i} H_{[\delta'+\gamma]}(X^{n_i}) \le   ( 1- \delta')(\overline{H}^*(\vect{X}) + \gamma). \nonumber 
\end{align}
Since $\gamma >0 $ is arbitrarily fixed and $H_{[\delta]}(X^{n})$ is a nonincreasing function of $\delta$, letting $\gamma \downarrow 0$, we obtain
\begin{align}
\liminf_{n \rightarrow \infty} \frac{1}{n} H_{[\delta]}(X^{n}) \le   ( 1- \delta')\overline{H}^*(\vect{X}). \label{eq:ineq0e}
\end{align}
Since $\delta' \in (0, \delta)$  is arbitrarily fixed, inequality \eqref{eq:ineq0e} implies $H_{[\delta]}^* (\vect{X}) \le (1-\delta) \overline{H}^*(\vect{X})$.

\smallskip
(iii) \emph{Proof of $H_{[\delta]} (\vect{X}) \le (1-\delta) \overline{H}(\vect{X})$}:~~This is a slightly strengthened version of the inequality given in \cite[Theorem 3]{Kuzuoka-Watanabe2015}, which demonstrates $\lim_{\gamma \downarrow 0} H_{[\delta + \gamma]} (\vect{X}) \le (1-\delta) \overline{H}(\vect{X})$.
This inequality can be proven similarly to case (ii).
\QED

\end{document}